\documentclass[a4paper,12pt]{article}
\newcommand{\doublespacing}{\let\CS=\@currsize\renewcommand{\baselinestretch}
{1.5}\tiny\CS}
\newcommand{\lyxaddress}[1]{
  \par {\raggedright #1
  \vspace{1.4em}
  \par}}

\begin{document}
\title{\textbf{Hardy's non-locality and generalized  non-local theory}}

\author{Sujit K. Choudhary \protect\( ^{1}\protect \)\
\thanks{sujit@imsc.res.in} ,
Sibasish Ghosh\protect\( ^{1}\protect \)\
\thanks{sibasish@imsc.res.in} , Guruprasad Kar\protect\( ^{2}\protect \)\
\thanks{gkar@isical.ac.in} , \\ Samir Kunkri\protect\( ^{3}\protect \)\
\thanks{skunkri@yahoo.com} , \ Ramij Rahaman\protect\(
^{2,4}\protect \)\
\thanks{Ramij.Rahaman@ii.uib.no} , \  Anirban Roy\protect\( ^{5}\protect \)\
\thanks{ranirb@gmail.com}}
\date{}
\maketitle

{ \lyxaddress{\protect\( ^{1}\protect \)The Institute of
Mathematical Sciences, C. I. T. Campus, Taramani, Chennai 600113,
India.}}

{ \lyxaddress{\protect\(^{2}\protect \)Physics and Applied
Mathematics Unit, Indian Statistical Institute, 203 B.T. Road,
Kolkata -700 108, India.}}

{ \lyxaddress{\protect\(^{3}\protect \)Mahadevananda Mahavidyalaya,
Monirampore, Barrackpore, North 24 Parganas, 700120, West Bengal,
India.}}

{ \lyxaddress{\protect\(^{4}\protect \)Selmer Center, Department of
Informatics, University of Bergen, Bergen-5020, Norway.}}

{ \lyxaddress{\protect\(^{5}\protect \)Department of Physics, Assam
 University Diphu Campus, Diphu 782462, Assam, India.}}

\maketitle

\begin{center}
{\bf Abstract}
\end{center}

{\small{Hardy's non-locality theorem for multiple two-level systems is explored in the context of
generalized non-local theory. We find non-local but non-signaling probabilities satisfying
Hardy's argument for two two-level and three two-level systems. Maximum probability
of success of Hardy's argument is obtained for three two-level systems in quantum theory
as well as in a more generalized theory. Interestingly, the maximum in the generalized
non-local theory for both the two two-level systems and three two-level systems turns
out to be the same.}

\section{Introduction}

There exist correlations between quantum systems which no local
realistic theory can reproduce. This was first shown by Bell by
means of an inequality, known as Bell's
inequality \cite{Bell}. Later, Hardy \cite{Hardy} gave an argument
which also reveals the non-local character of Quantum Mechanics.
His argument, unlike Bell's argument, does not use statistical inequalities
involving expectation values. This caused much interest among physicists.
The logical structure introduced by Hardy is as follows: Consider four yes-no
type events $A$, $B$, $A^{'}$ and $B^{'}$ where $A$ and  $A^{'}$
may happen in one system and $B$ and $B^{'}$ happen in another
system which is far apart from the first. The probability of joint
occurrence of $A^{'}$ and $B^{'}$  ({\it i.e.}, the joint
probability that both the events $A^{'}$ and $B^{'}$ are `yes') is
non-zero, $B^{'}$ always implies $A$ ({\it i.e.}, if $B^{'}$ is
`yes' then $A$ is also `yes'), $A^{'}$ always implies $B$ ({\it
i.e.}, if $A^{'}$ is `yes' then $B$ is also `yes'), but $A$ and
$B$ never occurs ({\it i.e.}, the joint probability that both $A$
as well as $B$ are `yes' is zero). These four statements are not
compatible with local realism. The nonzero probability appearing
in the argument is the measure of violation of local-realism. Any
given non-maximally entangled (pure) state of two two-level quantum systems
(\textit{i.e}., a system of two-qubits) exhibits Hardy's non-locality for
proper choices of observables but surprisingly no maximally
entangled state of such a system can show Hardy's non-locality.\\

Although no local-realistic theory can reproduce quantum
correlations still these correlations cannot be exploited to
communicate with a speed greater than that of the light in vacuum.
But quantum theory is not the only nonlocal theory consistent with
the relativistic causality \cite{popescu}. Theories which predict
nonlocal correlations and hence permit violation of Bell's
inequality but are constrained with the no signalling condition
are called `Generalized nonlocal theory(\textbf{GNLT})'. In recent
years there has been an increasing interest in \textbf{GNLT}
\cite {scarani, Barrett, cerf,
Acin, barrett, JBarret, Dam,Wolf, short, TShort}. In general, quantum
theory has been studied in the background of classical theory
which is comparatively restrictive. The new idea is to \emph{study
quantum theory from outside} i.e., starting from a general
family of theories, the so called `Generalized nonlocal
theory(\textbf{GNLT})' and to study properties common to all
\cite{Acin}. This might help in a better understanding of quantum
nonlocality.\\

In this paper, we study the Hardy's nonlocality argument in the
framework of \textbf{GNLT}. The maximum probability of success of
the Hardy's non-locality argument for two two-level quantum
systems is known to be 0.09 (approx) \cite{hardy93, jordan}. We find here
that if, instead of Quantum Mechanics, \textbf{GNLT} is
considered, the success probability of Hardy's argument can be
increased up to $50\%$ for two two-level systems.
For three two-level quantum systems, Wu and Xi
\cite{WX (1996)} have shown that any given genuinely entangled
pure state \footnote{pure states which are neither product states of three qubits nor a
product of one qubit state with an entangled state of the other two qubits
} of such systems will exihibit Hardy's nonlcality
provided the state satisfies a particular condition.
It turns out that almost all genuinely entangled three-qubit
pure states satisfy this particular condition. In a subsequent
development, Ghosh \emph{et al.}\cite{sghosh98} proved that Hardy's
nonlocality argument runs for any genuinely
entangled pure state of three qubits.
In particular, they showed that, in sharp contrast to bipartite cases,
every maximally entangled state of three qubits exhibits Hardy's
nonlocality and for each of these states, probability of success of Hardy's
argument can go maximum upto $12.5 \%$.  But, whether this is the maximum over all
pure entangled states of three qubits is worth searching. We find here
that this indeed is the case, that is, for three-qubit systems,
the Hardy's nonlocality argument can run maximum upto $12.5\%$.
We also study here the same in the context of \textbf{GNLT} and find that the
maximum probability reaches $50 \%$ which is
surprisingly the same as that for two two-level systems.\\

The paper is organized as follows. Section 2 deals with Hardy's
nonlocality argument for three two-level quantum systems. Hardy's argument for a general three
two-level systems in the framework of \textbf{GNLT} is discussed in Section 3.
For the sake of completeness of the argument, in Section 4, we descibe briefly the characteristic of no signalling Hardy type joint probabilities for a two two-level system in the framework of \textbf{GNLT}. Section 5 contains concluding remarks and some open questions. In Appendix I, we discribe the result of Section 4 in details. Appendix II describes the general scenario of Hardy's nonlocality argument for $n$ qubits.

\section{Hardy's non-locality for three qubit systems}

Let $|\psi\rangle$ be a pure state of three qubits 1,
2 and 3 jointly and let $(\hat{U}_j, \hat{D}_j)$ be a pair of $\{+1, -1\}$-valued
non-commuting observables for the $j$-th qubit (where
$j$ = 1, 2, 3). Considering here both $\hat{U}_j$ and $\hat{D}_j$ as $\{+1, -1\}$-valued
random variables associated to the $j$-th system (for $j=1,2,3$), the quantity
${\rm Prob} (R_1 = s_1, R_2 = s_2, R_3 = s_3)$ corresponds to the joint probability that the $3$-tuple
$(R_1,R_2,R_3) \in \{U_1, D_1\} \times \{U_2, D_2\} \times \{U_3, D_3\}$ of random variables $R_1, R_2$ and $R_3$ got the value $(s_1, s_2, s_3) \in \{+1, -1\}$. Thus, in the case of the three-qubit system $1+2+3$, the joint probability ${\rm Prob} (R_1 = s_1, R_2 = s_2, R_3 = s_3)$ would correspond to the joint probability
${\rm Prob} (\hat R_{1 }= s_1, \hat R_{2} = s_2, \hat R_{3} = s_3|\sigma_{123}) = \langle \hat R_{1 }= s_1, \hat R_{2} = s_2, \hat R_{3} = s_3|\sigma_{123}) |\hat R_{1 }= s_1, \hat R_{2} = s_2, \hat R_{3} = s_3\rangle$ for any density matrix $ \sigma_{123}$
of the three qubits $1, 2$ and $3$.
Note that here $\hat R_{j} |\hat R_{j}= s_j\rangle =s_j |\hat R_{j}= s_j\rangle$ for all $j= 1,2,3$
and $|\hat R_{1 }= s_1, \hat R_{2} = s_2, \hat R_{3} = s_3\rangle = |\hat R_{1 }= s_1\rangle \otimes |\hat R_{2} = s_2\rangle \otimes |\hat R_{3} = s_3\rangle$. Let us consider the following set of Hardy's nonlocality conditions
which is incompatible with local-realism \cite{sghosh98, kunkri}:

\begin{equation}
\label{hardynon} \begin{array}{lcl} {\rm Prob} (D_1 = + 1,
U_2 = + 1, U_3 = + 1) &=& 0,\\
{\rm Prob} (U_1 = + 1, D_2 = + 1, U_3 = + 1) &=& 0,\\
{\rm Prob} (U_1 = + 1, U_2 = + 1, D_3 = + 1) &=& 0,\\
{\rm Prob} (D_1 = - 1, D_2 = - 1, D_3 = - 1) &=& 0,\\
{\rm Prob} (U_1 = + 1, U_2 = + 1, U_3 = + 1) &>& 0.
\end{array}
\end{equation}

We first find out the set of all such states $|\psi\rangle$ of a three-qubit
system for which the probability ${\rm Prob}
(\hat{U}_1 = + 1, \hat{U}_2 = + 1, \hat{U}_3 = + 1| |\psi \rangle \langle \psi|)$= $|\langle \psi |
\hat{U}_1 = + 1, \hat{U}_2 = + 1, \hat{U}_3 = + 1 \rangle|^2$  is maximum for
a given set of observables $(\hat{U}_j,
\hat{D}_j)$ ($j = 1, 2, 3$). We would then maximize this maximum
probability by varying over the choice of the set of observables
$(\hat{U}_j, \hat{D}_j)$ ($j = 1, 2, 3$).

For this, let us take (for $j = 1, 2, 3$):
\begin{equation}
\label{basischoice}
\begin{array}{lcl}
|\hat{U}_j = + 1\rangle &=& a_j|\hat{D}_j = + 1\rangle +
b_j|\hat{D}_j
= - 1\rangle,\\
|\hat{U}_j = - 1\rangle &=& b_j^*|\hat{D}_j = + 1\rangle -
a_j^*|\hat{D}_j = - 1\rangle,
\end{array}
\end{equation}
where
\begin{equation}
\label{normalization1} |a_j|^2 + |b_j|^2 = 1~~ {\rm for}~ j = 1,
2, 3,
\end{equation}
and
\begin{equation}
\label{noncommuting} 0 < |a_j|, |b_j| < 1~~ {\rm for}~ j = 1, 2,
3.
\end{equation}

The last condition (given in equation (\ref{noncommuting})) is due
to the non-commutativity of $\hat{D}_j$ and $\hat{U}_j$ for each
$j = 1, 2, 3$. Let us first look for all those product states
$|{\psi}_p\rangle = |\phi\rangle \otimes |\eta\rangle \otimes
|\chi\rangle$ of the three-qubit system, each of which is orthogonal to
all the four product states: $|{\phi}_1\rangle \equiv |
\hat{D}_1 = + 1, \hat{U}_2 = + 1, \hat{U}_3 = + 1 \rangle$,
$|{\phi}_2\rangle \equiv | \hat{U}_1 = + 1, \hat{D}_2 = + 1,
\hat{U}_3 = + 1 \rangle$, $|{\phi}_3\rangle \equiv | \hat{U}_1 = +
1, \hat{U}_2 = + 1, \hat{D}_3 = + 1 \rangle$ and $|{\phi}_-\rangle
\equiv | \hat{D}_1 = - 1, \hat{D}_2 = - 1, \hat{D}_3 = - 1
\rangle$, appeared in the first four conditions of equation
(\ref{hardynon})\footnote{The motivation behind this is to identify all
the genuine three-qubit entangled states $|\psi\rangle$, each
of which satisfies the first four conditions given in equation (\ref{hardynon}) .}.
We must have therefore:
\begin{equation}
\label{productcondition}
\begin{array}{lcl}
\langle \phi | \hat{D}_1 = + 1 \rangle \langle \eta | \hat{U}_2 =
+
1 \rangle \langle \chi | \hat{U}_3 = + 1 \rangle &=& 0,\\
\langle \phi | \hat{U}_1 = + 1 \rangle \langle \eta | \hat{D}_2 =
+
1 \rangle \langle \chi | \hat{U}_3 = + 1 \rangle &=& 0,\\
\langle \phi | \hat{U}_1 = + 1 \rangle \langle \eta | \hat{U}_2 =
+
1 \rangle \langle \chi | \hat{D}_3 = + 1 \rangle &=& 0,\\
\langle \phi | \hat{D}_1 = - 1 \rangle \langle \eta | \hat{D}_2 =
- 1 \rangle \langle \chi | \hat{D}_3 = - 1 \rangle &=& 0.
\end{array}
\end{equation}

Choosing over all the different possibilities, that can satisfy
equation (\ref{productcondition}) and keeping in mind the
conditions (\ref{normalization1}) as well as (\ref{noncommuting}),
it can be shown that $|{\psi}_p\rangle$ can only be one of the
following three product states:
\begin{equation}
\label{productsolution}
\begin{array}{lcl}
|{\psi}_3\rangle &=& |\hat{U}_1 = - 1\rangle \otimes |\hat{U}_2 =
-
1\rangle \otimes |\hat{D}_3 = + 1\rangle,\\
|{\psi}_1\rangle &=& |\hat{D}_1 = + 1\rangle \otimes |\hat{U}_2 =
- 1\rangle \otimes |\hat{U}_3 = -
1\rangle,\\
|{\psi}_2\rangle &=& |\hat{U}_1 = - 1\rangle \otimes |\hat{D}_2 =
+ 1\rangle \otimes |\hat{U}_3 = - 1\rangle.
\end{array}
\end{equation}

Let ${\cal S}_1$ be the subspace (of the eight dimensional Hilbert
space ${\cal H}_{123}$ of the three qubits 1, 2, 3) spanned by the
product states $|{\phi}_1\rangle$, $|{\phi}_2\rangle$,
$|{\phi}_3\rangle$ and $|{\phi}_-\rangle$, and let ${\cal S}_2$ be
the subspace (of ${\cal H}_{123}$) spanned by the product states
$|{\psi}_1\rangle$, $|{\psi}_2\rangle$ and $|{\psi}_3\rangle$; then
${\cal S}_1$ must be orthogonal to ${\cal S}_2$. It can easily be
shown that the dimension of ${\cal S}_1$ is four whereas that of
${\cal S}_2$ is three. Thus the subspace ${\cal S}_1 \bigoplus
{\cal S}_2$ (of ${\cal H}_{123}$) has dimension seven. Therefore,
its orthogonal subspace $({\cal S}_1 \bigoplus {\cal S}_2)^{\bot}$
has to be one dimensional, and as $|{\psi}_1\rangle$, $|{\psi}_2\rangle$, $|{\psi}_3\rangle$
are the only three product states orthogonal to ${\cal S}_1$, $({\cal S}_1 \bigoplus {\cal S}_2)^{\bot}$ must
be spanned by a single entangled state $|{\psi}_0\rangle$ (say) of
the three-qubits. Thus it turns out that for $|\psi\rangle =
|{\psi}_0\rangle$, the probability $|\langle \psi | \hat{U}_1 = +
1, \hat{U}_2 = + 1, \hat{U}_3 = + 1 \rangle|^2$, appeared in the
last condition of equation (\ref{hardynon}), is maximum. It is noteworthy
that the state $|\hat{U}_1 = + 1, \hat{U}_2 = + 1,
\hat{U}_3 = + 1\rangle$ must be of the form:\footnote{As $|\hat{U}_1 = + 1, \hat{U}_2 = + 1,
\hat{U}_3 = + 1\rangle$ is linearly independent of (but not orthogonal to) $|{\phi}_1\rangle$, $|{\phi}_2\rangle$, $|{\phi}_3\rangle$, $|{\phi}_-\rangle$ and is orthogonal to $|{\psi}_1\rangle$, $|{\psi}_2\rangle$, $|{\psi}_3\rangle$.
}
$$|\hat{U}_1 = + 1, \hat{U}_2 = + 1, \hat{U}_3 =
+ 1\rangle = a|{\psi}_0\rangle + b|{\phi}_1\rangle +
c|{\phi}_2\rangle + d|{\phi}_3\rangle + e|{\phi}_-\rangle,$$ with $a\neq0$.

As $|{\psi}_0\rangle$ has to be orthogonal to all the linearly
independent states $|{\phi}_1\rangle$, $|{\phi}_2\rangle$,
$|{\phi}_3\rangle$, $|{\phi}_-\rangle$, $|{\psi}_1\rangle$,
$|{\psi}_2\rangle$ and $|{\psi}_3\rangle$, therefore
$|{\psi}_0\rangle$ must be unique. Taking this into account, it
can be shown that
$$|{\psi}_0\rangle =
|b_1b_2b_3|{\sqrt{\frac{1 - N}{N}}}\left[|\hat{U}_1 = + 1,
\hat{U}_2 = + 1, \hat{U}_3 = + 1\rangle -
\frac{a_3}{b_3^*}|\hat{U}_1 = + 1, \hat{U}_2 = + 1, \hat{U}_3 = -
1\rangle\right.$$
$$- \frac{a_2}{b_2^*}|\hat{U}_1 = +
1, \hat{U}_2 = - 1, \hat{U}_3 = + 1\rangle -
\frac{a_2a_3}{b_2^*b_3^*}\left({\frac{N}{1 - N}}\right)|\hat{U}_1
= + 1, \hat{U}_2 = - 1, \hat{U}_3 = - 1\rangle$$
$$- \frac{a_1}{b_1^*}|\hat{U}_1 = - 1, \hat{U}_2 = + 1, \hat{U}_3 = +
1\rangle - \frac{a_1a_3}{b_1^*b_3^*}\left({\frac{N}{1 -
N}}\right)|\hat{U}_1 = - 1, \hat{U}_2 = + 1, \hat{U}_3 = -
1\rangle$$
\begin{equation}
\label{psi0} \left.- \frac{a_1a_2}{b_1^*b_2^*}\left({\frac{N}{1 -
N}}\right)|\hat{U}_1 = - 1, \hat{U}_2 = - 1, \hat{U}_3 = +
1\rangle + \frac{a_1a_2a_3}{b_1^*b_2^*b_3^*}\left({\frac{N}{1 -
N}}\right)|\hat{U}_1 = - 1, \hat{U}_2 = - 1, \hat{U}_3 = -
1\rangle\right],
\end{equation}
where
\begin{equation}
\label{Nvalue} N = |b_1b_2|^2 + |b_2b_3|^2 + |b_3b_1|^2 -
2|b_1b_2b_3|^2 = |b_1b_2a_3|^2 + |a_1b_2b_3|^2 + |b_3b_1|^2.
\end{equation}

So, for given set of pairwise non-commuting observables
$(\hat{U}_j, \hat{D}_j)$ ($j = 1, 2, 3$), the maximum probability
$p(\hat{U}_1, \hat{D}_1, \hat{U}_2, \hat{D}_2, \hat{U}_3,
\hat{D}_3)$ (say), with which the Hardy-type non-locality
(\ref{hardynon}) will be satisfied, is given by
\begin{equation}
\label{observabledepprob} p(\hat{U}_1, \hat{D}_1, \hat{U}_2,
\hat{D}_2, \hat{U}_3, \hat{D}_3) = |\langle {\psi}_0 | \hat{U}_1 =
+ 1, \hat{U}_2 = + 1, \hat{U}_3 = + 1 \rangle|^2 =
|b_1b_2b_3|^2\left({\frac{1 - N}{N}}\right),
\end{equation}
where
\begin{equation}
\label{b's} |b_j|^2 = |\langle \hat{D}_j = - 1 | \hat{U}_j = + 1
\rangle|^2 = |\langle \hat{D}_j = + 1 | \hat{U}_j = - 1
\rangle|^2~~ {\rm for}~ j = 1, 2, 3,
\end{equation}
and $N$ is given by equation (\ref{Nvalue}). Any density matrix on
${\cal H}_{123}$, whose support is contained in $({\cal S}_1
\bigoplus {\cal S}_2)^{\bot} \bigoplus {\cal S}_2$ and which is
non-orthogonal to $|{\psi}_0\rangle$ will satisfy the Hardy-type
non-locality conditions given in equation (\ref{hardynon}), no
other density matrix of ${\cal H}_{123}$ can satisfy it.\\

We would now maximize the probability $p(\hat{U}_1, \hat{D}_1,
\hat{U}_2, \hat{D}_2, \hat{U}_3, \hat{D}_3)$ over all possible
choices of the pairwise non-commuting observables $(\hat{U}_j,
\hat{D}_j)$ ($j = 1, 2, 3$). From equations
(\ref{observabledepprob}) and (\ref{Nvalue}), $p$ is a
symmetric function of the three variables $|b_1|^2$, $|b_2|^2$ and
$|b_3|^2$, where $0 < |b_1|, |b_2|, |b_3| < 1$ and so it will attain
its extremum only when $|b_1|^2 = |b_2|^2 = |b_3|^2 = k$ (say).
Taking $|b_1|^2 = |b_2|^2 = |b_3|^2 = k$ in the expression for
$p$, we find that
\begin{equation}
\label{psymmvalue} p \equiv p(k) = \frac{k}{3 - 2k} - k^3,~~ {\rm
with}~ 0 < k < 1.
\end{equation}
So $p(k)$ attains its maximum value when $k = 1/2$. Thus we see
that the maximum value 1/8 of $p(\hat{U}_1, \hat{D}_1, \hat{U}_2,
\hat{D}_2, \hat{U}_3, \hat{D}_3)$ occurs when and only when
$|b_j|^2 = |\langle \hat{D}_j = - 1 | \hat{U}_j = + 1 \rangle|^2 =
|\langle \hat{D}_j = + 1 | \hat{U}_j = - 1 \rangle|^2 = 1/2$ for
$j = 1, 2, 3$. In this case, $|{\psi}_0\rangle$ is given by
$$|{\psi}_0\rangle = \frac{1}{\sqrt{2}}|\hat{U}_1 = + 1\rangle
\otimes \left\{\frac{1}{\sqrt{2}}|\hat{U}_2 = + 1\rangle \otimes
\frac{1}{\sqrt{2}}\left(|\hat{U}_3 = + 1\rangle - e^{i(x_3 +
y_3)}|\hat{U}_3 = - 1\rangle\right)\right.$$
$$\left.- \frac{e^{i(x_2 + y_2)}}{\sqrt{2}}|\hat{U}_2 = - 1\rangle \otimes
\frac{1}{\sqrt{2}}\left(|\hat{U}_3 = + 1\rangle + e^{i(x_3 +
y_3)}|\hat{U}_3 = - 1\rangle\right)\right\}$$
$$- \frac{e^{i(x_1 + y_1)}}{\sqrt{2}}|\hat{U}_1 = - 1\rangle \otimes
\left\{\frac{1}{\sqrt{2}}|\hat{U}_2 = + 1\rangle \otimes
\frac{1}{\sqrt{2}}\left(|\hat{U}_3 = + 1\rangle + e^{i(x_3 +
y_3)}|\hat{U}_3 = - 1\rangle\right)\right.$$
\begin{equation}
\label{psimax} \left. + \frac{e^{i(x_2 +
y_2)}}{\sqrt{2}}|\hat{U}_2 = - 1\rangle \otimes
\frac{1}{\sqrt{2}}\left(|\hat{U}_3 = + 1\rangle - e^{i(x_3 +
y_3)}|\hat{U}_3 = - 1\rangle\right)\right\},
\end{equation}
which is nothing but a maximally entangled state of three qubits
(and hence, it is local-unitarily connected to the three qubit
GHZ state for which the maximum success probability of Hardy's
argument is known to be $12.5\%$). Here $a_j = \langle \hat{D}_j =
+ 1 | \hat{U}_j = + 1 \rangle = - {\langle \hat{D}_j = - 1 |
\hat{U}_j = - 1 \rangle}^* = \frac{e^{ix_j}}{\sqrt{2}}$ and $b_j =
\langle \hat{D}_j = - 1 | \hat{U}_j = + 1 \rangle = {\langle
\hat{D}_j = + 1 | \hat{U}_j = - 1 \rangle}^* =
\frac{e^{iy_j}}{\sqrt{2}}$ for $j = 1, 2, 3$. Thus we conclude
this section with the fact that the maximum probability of success
of Hardy's argument for three-qubit states is $12.5\%$ and the
state which responds to this maximum is the three-qubit GHZ state (upto local unitary).\\

\section{General non-signaling probabilities satisfying Hardy-type
non-locality for three two-level systems}

In the framework of a general probabilistic theory, consider a physical
system consisting of three subsystems shared among three far apart
parties Alice, Bob and Charlie. Assume that Alice, Bob and Charlie
can measure one of the two observables $X_i$ and $Y_i$, where $i$
stands for the 1st ({\it i.e.}, Alice), 2nd ({\it i.e.}, Bob), or
3rd ({\it i.e.}, Charlie), on their respective subsystems. The
outcomes of each such measurement can be either `up'($+1$) or `down' ($-1$)
. We now consider all the sixty four joint probabilities
${\rm Prob} (R_1 = s_1, R_2 = s_2, R_3 = s_3)$,
where $R_i \in \{X_i, Y_i\}$ for $i=1,2,3$ and $s_1, s_2, s_3
\in \{+1, -1\}$. Let us now impose the normalization, non-signalling as well as
Hardy's non-locality type conditions on these sixty four joint
probabilities :\\

{\noindent {\bf{Condition (1): ({\it Normalization conditions}):}}}\\
\begin{equation}
\label{normalization}
\begin{array}{lcl}
\sum_{s_1,s_2,s_3 \in \{+1,-1\} }{\rm Prob} (R_1 = s_1, R_2 = s_2, R_3 = s_3)=1 ~{\rm for~all}\\\\
(R_1,R_2,R_3)\in\{X_1,Y_1\}\times \{X_2,Y_2\} \times \{X_3,Y_3\}.
\end{array}
\end{equation}

{\noindent {\bf Condition (2): ({\it Non-signalling conditions}):}}\\
For the marginal joint probabilities of the first and second particles, we must have
\begin{equation}
\label{nsforcharlie}
\begin{array}{lcl}
\sum_{s_3 \in \{+1,-1\} }{\rm Prob} (R_1 = s_1, R_2 = s_2, X_3 = s_3)=\sum_{s_3 \in \{U,D\} }{\rm Prob} (R_1 = s_1, R_2 = s_2, Y_3 = s_3)\\\\
 ~{\rm for~all}~(R_1,R_2)\in\{X_1,Y_1\}\times \{X_2,Y_2\} ~{\rm and}~{\rm for~all}~ s_1,s_2 \in \{+1,-1\},\\\\
{\rm and~ similarly~ for~ the~ marginal~ probabilities~ of~ the~ first~ and~ third~ parties~} \\{\rm   as~ well~ as~
~for~the~ second~ and~ third~ parties,~ separately.}
\end{array}
\end{equation}

{\noindent {\bf Condition (3): ({\it Hardy-type
non-locality
conditions}):}}\\
\begin{equation}
\label{hardy}
\begin{array}{lcl}
{\rm Prob}(Y_1 = +1, X_2 = +1, X_3 = +1)= 0,\\
{\rm Prob}(X_1 = +1, Y_2 = +1, X_3 = +1)= 0,\\
{\rm Prob}(X_1 = +1, X_2 = +1, Y_3 = +1)= 0,\\
{\rm Prob}(Y_1 = -1, Y_2 = -1, Y_3 = -1)= 0,\\
{\rm Prob}(X_1 = +1, X_2 = +1, X_3 = +1) > 0.
\end{array}
\end{equation}
\vspace{0.4cm}
Our aim is to maximize the probability ${\rm Prob}(X_1 = +1, X_2 = +1, X_3 = +1)$
subject to satisfying all the conditions given in equations
(\ref{normalization}), (\ref{nsforcharlie}) (\ref{hardy}).\\
We have maximized for ${\rm Prob}(X_1 = +1, X_2 = +1, X_3 = +1)$ with the help of
\emph{Mathematica} and have found this maximum value, ${\rm Prob}^{max}(X_1 = +1, X_2 = +1, X_3 = +1)$ to be $0.5$,
while the rest of the sixty four probabilities are given by
\footnote{This solution set is not unique, but ${\rm Prob^{max}}(X_1 = +1, X_2 = +1, X_3 = +1) = 0.5$ for each such set.}
$$
{\rm Prob}(X_1 = +1, X_2 = +1, X_3 = +1)=\\
{\rm Prob}(X_1 = +1, X_2 = +1, Y_3 = -1)= 0.5,$$
$${\rm Prob}(X_1 = +1, Y_2 = -1, X_3 = +1)= \\
{\rm Prob}(X_1 = +1, Y_2 = -1, Y_3 = -1)= 0.5,$$
$${\rm Prob}(X_1 = -1, X_2 = +1, X_3 = -1)=\\
{\rm Prob}(X_1 = -1, X_2 = +1, Y_3 = +1)= 0.5,$$
$${\rm Prob}(X_1 = -1, Y_2 = -1, X_3 = -1)=\\
{\rm Prob}(X_1 = -1, Y_2 = -1, Y_3 = +1)= 0.5,$$
$${\rm Prob}(Y_1 = +1, X_2 = +1, X_3 = -1)= \\
{\rm Prob}(Y_1 = +1, X_2 = +1, Y_3 = -1)= 0.5,$$
$${\rm Prob}(Y_1 = +1, Y_2 = -1, X_3 = -1)= \\
{\rm Prob}(Y_1 = +1, Y_2 = -1, Y_3 = -1)= 0.5,$$
$${\rm Prob}(Y_1 = -1, X_2 = +1, X_3 = +1)= \\
{\rm Prob}(Y_1 = -1, X_2 = +1, Y_3 = +1)= 0.5,$$
$${\rm Prob}(Y_1 = -1, Y_2 = -1, X_3 = +1)= \\
{\rm Prob}(Y_1 = -1, Y_2 = -1, Y_3 = +1)= 0.5,$$
${\rm with~ the~ remaining~ probabilities~ being~ all~ zero.}
$\\

Thus the maximum probability of success of Hardy's argument for
three two-level systems is more in GNLT than in the quantum
theory.

\section{General non-signaling probabilities satisfying Hardy-type
non-locality for two two-level systems}

In this section, we will descibe briefly the character of a set of sixteen non-signaling
joint probabilities ${\rm Prob} (M = m, N = n)$, where $m, n \in
\{+ 1, - 1\}$, $M$ is one of the two $\{+ 1, - 1\}$-valued random
observables $A$, $A'$, chosen by Alice, and $N$ is one of the two
$\{+ 1, - 1\}$-valued random observables $B$, $B'$, chosen by
Bob. Other than being members of the interval $[0, 1]$, these sixteen
probabilities must satisfy the normalization conditions. This will lead
us (conditions (\ref{normalization1}) in Appendix I) to a set of four equations.
The no-signaling constraint ({\it i.e.}, the relativistic
causality) results in a set of eight linear equations
(conditions (\ref{nsforcharlie1})-(\ref{nsforcharlie2}) in Appendix I).
If we further impose the restriction that four of these
probabilities respect Hardy's non-locality conditions (conditions (\ref{hardycondition2q}) in Appendix I), it can
be easily seen (see Appendix I)  that the nonzero probability appearing in Hardy's argument can
at most go to $\frac{1}{2}$ and for the nonzero probability appearing in Hardy's
argument equals to $\frac{1}{2}$,  there is a {\it unique}
solution for the above-mentioned sixteen joint probabilities
satisfying simultaneously all the above mentioned conditions viz.
the normalization condition, the causality condition and the Hardy's condition.\\

Thus we see that above-mentioned sixteen probabilities will be
{\it non-local} as well as {\it non-signaling} if the probability of success of
Hardy's argument lies between $0$ and
$1/2$. We know that quantum states cannot exhibit Hardy's
nonlocality with a probability more than 0.09 but in a more
generalized nonlocal theory, the success probability of Hardy's
argument can be increased up to 0.5.\\

These sixteen probabilities give rise to a unit positive cube in sixteen dimensional Euclidean space. There are, in all, twelve constraints due to the normalization and the causality conditions. But not all these constraints are linearly independent, only eight of them are so. So the solution space is given by an eight dimensional polytope inscribed within this unit cube. This polytope is called the 'Causal polytope' $\it{\textbf{C}}$. It has been found that there are 24 vertices of this polytope \cite{barrett}, 16 of which represent local correlations (which form the `local polytope' $\textbf{L}$ or 'Bell polytope') and 8 represent nonlocal correlations.  The Bell polytope lies inside the causal polytope. Quantum correlations can lie outside the Bell polytope but are always inside $\textbf{C}$. They are further restricted by the Cirelson's bound. The maximum algebraic value of the Bell-CHSH expression is 4 \cite{popescu} which corresponds to the nonlocal vertices of $\textbf{C}$. Imposition of Hardy's condition(\ref{hardycondition2q}) (in addition to the normalization and the causality conditions) restricts the solution space to a four dimensional subspace of the polytope $\textbf{C}$ called the Hardy's polytope $\textbf{H}$. Quantum correlations can lie outside this polytope but are always inside $\textbf{C}$ and are further restricted by the value $0.09$ for the success probability of Hardy's argument. As shown above in \textbf{GNLT}, the success probability of Hardy's argument can go maximum upto $0.5$ which also corresponds to the nonlocal vertices of the causal polytope\footnote{Each nonlocal vertex  corresponds to a different nonlocal non-quantum Hardy type condition.}

\section{Conclusion}

In conclusion, we have shown here that for three two level quantum
systems the maximum probability of success of Hardy's argument is
$12.5\%$ which in fact is exhibited by the maximally entangled
state of three qubits namely the GHZ state. The corresponding
maximum probability for two two-level quantum system is known to
be $9\%$ (approx.).It would have been interesting to search whether this
trend is sustained or not \emph{i.e.}, whether the success
probability of Hardy's argument keeps on increasing with increase
in number of qubits or not as it is known in this context that
violation of Bell's inequality increases exponentially with the
increase in number of qubits which falls highly against the common
acceptance \cite{N. D. Mermin, S M Roy, M.Ardehali, Klyscho, S. L.
Braunstein, Adan Cabello}. For this, in the Appendix II, we have
provided a way to find out the state of $n$ two-level quantum
systems which will show maximum departure from local-realism if
Hardy's logical structure is opted for this purpose.\\

If instead of quantum mechanics a more general framework
of GNLT is adopted, the maximum probability of success of Hardy's
argument can be enhanced for both the three two-level systems and
for two two-level systems. Interestingly, the maximum success
probability for both type of systems attains a common value
$0.5$.\\

Quantum non-locality has attracted much attention since its
discovery because it relates quantum mechanics with special
relativity. Special relativity forbids sending physical
information with a speed greater than that of the light in vacuum.
This is reflected in the quantum mechanical joint probabilities
appearing both in the violation of Bell's inequality as well as in
the fulfillment of Hardy's non-locality conditions, although there
is no direct relevance of special theory of relativity in the
postulates of non-relativistic quantum mechanics. These quantum
mechanical joint probabilities are not only non-local but also
non-signaling. The non-local probabilities, coming out from
quantum mechanical states can give rise to the maximum violation
up to the amount $2{\sqrt{2}}$ of Bell's inequality, whereas there
are non-quantum mechanical non-local joint probabilities which
give rise to the maximal possible algebraic violation (namely,
$4$) of Bell's inequality, without violating the
relativistic-causality \cite{popescu}. Why quantum theory can not
provide more than $2{\sqrt{2}}$ violation of the Bell's
inequality? By exploiting the theoretical structure of quantum
mechanics it has been shown that a violation greater than
$2{\sqrt{2}}$ will result in signalling in quantum mechanics
\cite{Dieks02,Buhr,sk}. We have seen in this paper that the
no-signaling constraint cannot restrict the maximum value of the
non-zero probability appearing in the Hardy's argument to $0.09$
all by itself. In a generalized non-signalling theory this value
can go up to $0.5$. It will be an interesting open question to
find what feature of quantum mechanics along with no-signalling
condition restricts the value to
$0.09$\\

\section{Acknowledgments}
 R. R. acknowledges the partial supports by Norwegian Research Council (Norway) as well
as by the CSIR, Govt. of India (during his stay
at Indian Statistical Institute, Kolkata). A major part of this work was done when
S. K. C. and R. R. were in Physics and Applied Mathematics Unit, Indian Statistical Institute, Kolkata, India. The authors gratefully acknowledge the anonymous referee for giving
suggestions to revise the earlier version of the manuscript which resulted in its present form.\\

\newpage
\begin{center}
\textbf{\large{Appendix I}}
\end{center}
The sixteen joint probabilities mentioned in section 4 of the paper are of the form:
$$
\begin{array}{lcl}
{\rm Prob} (R_1 = s_1, R_2 = s_2) ~{\rm for~all}
~(R_1,R_2)\in\{A,A'\}\times \{B,B'\}\\ {\rm and~for~all}~ (s_1,s_2)\in\{+1, -1\}.
\end{array}
$$
The normalization conditions on these joint probabilities are given by :
\begin{equation}
\label{normalization1}
\begin{array}{lcl}
\sum_{s_1,s_2} \in \{+1,-1\} {\rm Prob} (R_1 = s_1, R_2 = s_2) = 1 ~{\rm for~all}\\\\
(R_1,R_2)\in\{A, A'\}\times \{B, B'\}.
\end{array}
\end{equation}
Now the no-signaling constraint ({\it i.e.}, the relativistic
causality) implies that if Alice performs the experiment for $A$
(or $A^{'}$), the individual probabilities for the outcomes $A = +
1$ (or $A^{'} = + 1$) and $A = - 1$ (or $A^{'} = - 1$) must be
independent of whether Bob chooses to perform the experiment for
$B$ or $B^{'}$ and similar should be the case for Bob also. So for
the above-mentioned sixteen probabilities, the condition for
causality to hold is given by:
\begin{equation}
\label{nsforcharlie1}
\begin{array}{lcl}
\sum_{s_2 \in \{+1,-1\} }{\rm Prob} (R_1 = s_1, B = s_2)=\sum_{s_2 \in \{+1, -1\} }{\rm Prob} (R_1 = s_1, B' = s_2)\\\\
 ~{\rm for~all}~R_1\in\{A ,A'\}~{\rm and}~{\rm for~all}~ s_1 \in \{+1,-1\},\\\\
 \end{array}
\end{equation}
and
\begin{equation}
\label{nsforcharlie2}
\begin{array}{lcl}
\sum_{s_1 \in \{+1,-1\} }{\rm Prob} (A = s_1, R_2 = s_2)=\sum_{s_1 \in \{+1, -1\} }{\rm Prob} (A' = s_1, R_2 = s_2)\\\\
 ~{\rm for~all}~R_2\in\{B, B'\}~{\rm and}~{\rm for~all}~ s_2 \in \{+1,-1\}
\end{array}
\end{equation}

We further assume that the above-mentioned probabilities respect
the Hardy's non-locality conditions\footnote{The conditions
given in equation (19) are not compatible with the notion of
local-realism. To see this let us consider those local-realistic
states for which values of both $A'$ and $B'$ are $+1$. The
condition $q> 0$ guarantees that there exists such states. Now
for these states ${\rm Prob}(A = -1, B' = +1)= 0$ and ${\rm Prob}(A' = +1, B = -1)= 0$ respectively imply  $A=+1$ and
$B=+1$. But this contradicts the first condition ${\rm Prob}(A = +1, B = +1)=0$. }:
\begin{equation}
\label{hardycondition2q}
\begin{array}{lcl}
{\rm Prob}(A = +1, B = +1)= {\rm Prob}(A' = +1, B = -1)= {\rm Prob}(A = -1, B' = +1)= 0,\\
{\rm Prob}(A' = +1, B' = +1)= q> 0
\end{array}
\end{equation}

Using equation (\ref{hardycondition2q}) into equation
(\ref{nsforcharlie1}) corresponding to $R_1= A$ and $s_1= +1$, we get
\begin{equation}
\label{ineq2q1st}
{\rm Prob}(A = +1, B = -1) \ge {\rm Prob}(A' = +1, B' = +1).
\end{equation}
Using equation (\ref{hardycondition2q}) into equation
(\ref{nsforcharlie2}), corresponding to $R_2= B$ and $ s_2= +1$, we get
\begin{equation}
\label{ineq2q2nd}
{\rm Prob}(A = -1, B = +1) \ge {\rm Prob}(A' = +1, B = +1).
\end{equation}
Using equation (\ref{hardycondition2q}) into equation
(\ref{normalization1}) corresponding to $R_1= A$ and $R_2= B$, we get
\begin{equation}
\label{ineq2q3rd}
\begin{array}{lcl}
1 = {\rm Prob}(A = +1, B = -1)+{\rm Prob}(A = -1, B = +1)+{\rm Prob}(A = -1, B = -1)
 \ge \\
{\rm Prob}(A = +1, B = -1)+{\rm Prob}(A = -1, B = +1)
\ge\\
{\rm Prob}(A' = +1, B = +1)+{\rm Prob}(A = +1, B' = +1),
\end{array}
\end{equation}
using equations (\ref{ineq2q1st}) and (\ref{ineq2q2nd}). Using
equation (\ref{hardycondition2q}) into equations
(\ref{nsforcharlie1}) corresponding to $R_1=A$ and $s_1=+1$ and (\ref{nsforcharlie2}) corresponding
to $R_2=B'$ and $s_2=+1$ , we get
\begin{equation}
\label{ineq2q4th}
{\rm Prob}(A' = +1, B = +1)+ {\rm Prob}(A = +1, B' = +1) =$$$$ 2q+ {\rm Prob}(A' = +1, B' = -1)+{\rm Prob}(A' = -1, B' = +1)\ge 2q.
\end{equation}
Using equations (\ref{ineq2q3rd}) and (\ref{ineq2q4th}), we get
$$1 \ge {\rm Prob}(A' = +1, B = +1)+ {\rm Prob}(A = +1, B' = +1) \ge 2q.$$
Thus we have
\begin{equation}
\label{ineq2q5th} q \le \frac{1}{2}.
\end{equation}
If we now follow the argument, beginning at equation
(\ref{ineq2q1st}) and ending at equation (\ref{ineq2q5th}), it can
be easily shown that for $q = 1/2$, there is a {\it unique}
solution for the above-mentioned sixteen joint probabilities
satisfying simultaneously all the conditions
(\ref{normalization1}) to (\ref{hardycondition2q}):
$$
{\rm Prob}(A = +1, B = -1)= {\rm Prob}(A = -1, B = +1)= \\
{\rm Prob}(A' = +1, B = +1)=  0.5
$$
$${\rm Prob}(A' = -1, B = -1)= {\rm Prob}(A = +1, B' = +1)= \\
{\rm Prob}(A = -1, B' = -1)=  0.5
$$
$${\rm Prob}(A' = -1, B' = -1)=  0.5$$
and
$$
{\rm Prob}(A = -1, B = -1)= {\rm Prob}(A' = -1, B = +1)= 0 \\
$$
$$
{\rm Prob}(A = +1, B' = -1)= {\rm Prob}(A' = +1, B' = -1)= \\
{\rm Prob}(A' = -1, B' = +1)=  0
$$
Thus we see that above-mentioned sixteen probabilities will be
{\it non-local} as well as {\it non-signaling} iff  $0 < q \le
1/2$.
\newpage
\begin{center}
\textbf{\large{Appendix II}}
\end{center}
Here we try to find out the state of $n$ qubits
($n > 3$) which exhibits the Hardy's nonlocality conditions for $n$ two-level
systems with maximum nonlocal
probability. For the choice of $\{+1, -1\}$-valued two non-commuting
observables $\hat{U}_j$ and $\hat{D}_j$ for the $j$-th qubit,
Hardy's non-locality argument runs as follows (for an $n$-qubit state $|\psi\rangle$):
\begin{equation}
\label{n-hardy}
\begin{array}{lcl}
\left\langle\psi| \hat{D}_1 = + 1, \hat{U}_2 = + 1, \hat{U}_3 = + 1,
\ldots, \hat{U}_n = + 1 \right\rangle &=& 0,\\
\left\langle \psi|\hat{U}_1 = + 1, \hat{D}_2 = + 1, \hat{U}_3 = + 1,
\ldots, \hat{U}_n = + 1 \right\rangle &=& 0,\\
\ldots  \ldots & & \ldots,\\
\left\langle \psi|\hat{U}_1 = + 1, \hat{U}_2 = + 1, \ldots, \hat{U}_{n
-
1} = + 1, \hat{D}_n = + 1 \right\rangle &=& 0,\\
\left\langle \psi|\hat{D} = - 1, \hat{D}_2 = - 1, \hat{D}_3 = - 1,
\ldots, \hat{D}_n = - 1 \right\rangle &=& 0,\\
|\left\langle \psi|\hat{U}_1 = + 1, \hat{U}_2 = + 1, \hat{U}_3 = + 1,
\ldots, \hat{U}_n = + 1 \right\rangle|^2 &>& 0.
\end{array}
\end{equation}
As above, if $S_1$ is the $(n + 1)$-dimensional subspace of the
$n$-qubit Hilbert space $({C\!\!\!\!I}^2)^{\otimes n}$, linearly
spanned by the $(n + 1)$ number of linearly independent product
states $|\hat{D}_1 = + 1, \hat{U}_2 = + 1, \hat{U}_3 = + 1,
\ldots, \hat{U}_n = + 1\rangle$, $|\hat{U}_1 = + 1, \hat{D}_2 = +
1, \hat{U}_3 = + 1, \ldots, \hat{U}_n = + 1\rangle$, $\ldots$,
$|\hat{U}_1 = + 1, \hat{U}_2 = + 1, \ldots, \hat{U}_{n - 1} = + 1,
\hat{D}_n = + 1\rangle$, $|\hat{D}_1 = - 1, \hat{D}_2 = - 1,
\hat{D}_3 = - 1, \ldots, \hat{D}_n = - 1\rangle$, then all the
(fully) product states, each of which is orthogonal to $S_1$, will
linearly span the $(2^n - n - 2)$-dimensional subspace $S_2$. So
$S_2$ is orthogonal to $S_1$. One can also show that $S_2$ can be
linearly spanned by the following $(2^n - n - 2)$ number of
linearly independent product states:
$$\left|\hat{D}_1 = + 1, \hat{U}_2 = - 1, \hat{U}_3 = - 1, \ldots,
\hat{U}_n = - 1\right\rangle, \left|\hat{U}_1 = - 1, \hat{D}_2 = +
1, \hat{U}_3 = - 1, \ldots, \hat{U}_n = - 1\right\rangle,$$
$$\ldots, \left|\hat{U}_1 = - 1, \hat{U}_2 = - 1, \ldots, \hat{U}_{n - 1} = -
1, \hat{D}_n = + 1\right\rangle;~ ({\rm total}~ {\rm no.}~ = n)$$\\
$$\left|\hat{D}_1 = + 1, \hat{D}_2 = + 1, \hat{U}_3 = - 1, \ldots,
\hat{U}_n = - 1\right\rangle, \left|\hat{D}_1 = + 1, \hat{U}_2 = -
1, \hat{D}_3 = + 1, \ldots, \hat{U}_n = - 1\right\rangle,$$
$$\ldots, \left|\hat{U}_1 = - 1, \hat{U}_2 = - 1, \ldots, \hat{D}_{n
- 1} = + 1, \hat{D}_n = + 1\right\rangle;~ ({\rm total}~ {\rm
no.}~
= \frac{n(n - 1)}{2})$$\\
$$\ldots \ldots \ldots$$\\
$$\ldots \ldots \ldots$$\\
$$\left|\hat{D}_1 = + 1, \ldots, \hat{D}_{n - 2} = + 1, \hat{U}_{n -
1} = - 1, \hat{U}_n = - 1\right\rangle,$$
$$\left|\hat{D}_1 = + 1, \ldots, \hat{D}_{n - 3} = + 1, \hat{U}_{n - 2} = - 1, \hat{D}_{n -
1} = + 1, \hat{U}_n = - 1\right\rangle, \ldots,$$
\begin{equation}
\label{productlabels} \left|\hat{U}_1 = - 1, \hat{U}_2 = - 1,
\hat{D}_3 = + 1, \ldots, \hat{D}_n = + 1\right\rangle~ ({\rm
total}~ {\rm no.}~ = \frac{n(n - 1)}{2}).
\end{equation}
Note that the product state $|\hat{U}_1 = + 1, \hat{U}_2 = + 1,
\hat{U}_3 = + 1, \ldots, \hat{U}_n = + 1\rangle$ (which appeared
in the last condition in equation (\ref{n-hardy})), is orthogonal
to each of the $(2^n - n - 2)$ product states appeared in equation
(\ref{productlabels}), and hence, it is orthogonal to $S_2$ as
well. So, in order that the inequality in the last condition of
equation (\ref{n-hardy}) is satisfied, the state $|\hat{U}_1 = +
1, \hat{U}_2 = + 1, \hat{U}_3 = + 1, \ldots, \hat{U}_n = +
1\rangle$ has to have a non-zero overlap with the one-dimensional
subspace $(S_1 \bigoplus S_2)^{\bot}$ of
$({C\!\!\!\!I}^2)^{\otimes n}$. Let $|\psi_0\rangle$ be the
(entangled) state spanning the subspace $(S_1 \bigoplus
S_2)^{\bot}$. Thus we see that, for the given set of $\{+1, -1\}$-valued,
pairwise non-commuting observables $\{(\hat{U}_j, \hat{D}_j) | j =
1, 2, \ldots, n\}$, there exits a unique state $|\psi_0\rangle$
satisfying the Hardy's non-locality conditions (\ref{n-hardy})
with maximum non-local probability. This $|\psi_0\rangle$ can now be found
out easily as it is orthogonal to the $(2^n - 1)$ number of
linearly independent product states (described above) spanning
$(S_1 \bigoplus S_2)$. One can then also try to maximize the
probability $|\langle \psi_0 | \hat{U}_1 = + 1, \hat{U}_2 = + 1,
\ldots, \hat{U}_n = + 1 \rangle|^2 $ over all possible
choices of the set $\{(\hat{U}_j, \hat{D}_j) | j = 1, 2, \ldots,
n\}$ of observables (and thereby, over all possible choices of
$|\psi_0\rangle$).

\end{document}